Title: A Semantic Cross-Species Derived Data Management Application

Authors: David B. Keator[1], Jinran Chen[1], B Nolan Nichols[2], Fariba Fana[3], Hal Stern[4], Tallie Z. Baram[5], Steven L. Small[6].

Affiliations:
[1] Department of Psychiatry and Human Behavior, University of California, Irvine, CA., USA.
[2] Department of Bioinformatics and Computational Biology, Genentech, Inc., South San Francisco, CA., USA
[3] Qualcomm Institute, University of California, San Diego, CA., USA.
[4] Department of Statistics, University of California, Irvine, CA., USA.
[5] Departments of Pediatrics, Anatomy and Neurobiology, University of California, Irvine, CA., USA.
[6] Department of Neurology, University of California, Irvine, CA., USA.

Corresponding Author:

David B. Keator, Ph.D.
University of California, Irvine
Neuroscience Imaging Center
Irvine Hall rm. 163 - ZOT 3960
Irvine, CA. 92697
dbkeator@uci.edu

**Abstract:**
Managing dynamic information in large multi-site, multi-species, and multi-discipline consortia is a challenging task for data management applications. Often in academic research studies the goals for informatics teams are to build applications that provide extract-transform-load (ETL) functionality to archive and catalog source data that has been collected by the research teams. In consortia that cross species and methodological or scientific domains, building interfaces that supply data in a usable fashion and make intuitive sense to scientists from dramatically different backgrounds increases the complexity for developers. Further, reusing source data from outside one's scientific domain is fraught with ambiguities in understanding the data types, analysis methodologies, and how to combine the data with those from other research teams. We report on the design, implementation, and performance of a semantic data management application to support the NIMH funded Conte Center at the University of California, Irvine. The Center is testing a theory of the consequences of "fragmented" (unpredictable, high entropy) early-life experiences on adolescent cognitive and emotional outcomes in both humans and rodents. It employs cross-species neuroimaging, epigenomic, molecular, and neuroanatomical approaches in humans and rodents to assess the potential consequences of fragmented unpredictable experience on brain structure and circuitry. To address this multi-technology, multi-species approach, the system uses semantic web techniques based on the Neuroimaging Data Model (NIDM) to facilitate data ETL functionality. We find this approach enables a low-cost, easy to maintain, and semantically meaningful information management system, enabling the diverse research teams to access and use the data.

1. Introduction

Managing dynamic information in large multi-site, multi-species, and multi-discipline consortia is a challenging task for data management applications. Often in academic research studies the goals for informatics teams are to build applications that provide extract-transform-load (ETL) functionality to archive and catalog source data that has been collected by the research teams. In consortia that cross species and scientific domains, building interfaces that supply data in a reusable fashion and make intuitive sense to scientists from dramatically different backgrounds increases the complexity for both the users and developers. Further, reusing source data from outside one's scientific domain is fraught with ambiguities in understanding the data types, typical analysis methodologies, and ways of reusing the data and integrating them with those from other research teams to make the desired scientific inferences.

In this work, we develop a data management application to support geographically distributed, multi-disciplinary, teams for the Conte Center on Brain Programming in Adolescent Vulnerabilities at the University of California, Irvine (1P50MH096889-01A1)[1]. The Conte Center consists of four main research teams focused on studying the effect of early life "fragmented" (high entropy) early-life experiences, and especially maternal care patterns, on cognitive and emotional vulnerabilities in adolescence, across species[2]. Team one focuses on the mechanisms of emotional and cognitive vulnerabilities after fragmented early life in rodent models. Team two addresses fragmentation in the prenatal environment in humans, collecting measurements of fragmented and unpredictable shifts in maternal mood. Team three focuses on pre- and post-natal exposure to fragmented maternal signals during infancy, basing the analyses on structured sessions between mother and infant using audiovisual material, and collecting outcome measures during childhood and adolescence that include MRI, psychological and clinical interviews, and formal scales. Team four focuses on structural and functional network correlates of fragmented early life across species as measured with magnetic resonance imaging (MRI). The Center further consists of a neuroimaging core and a biostatistics, computational, and data management (BCDM) core that work in an integrated manner. A major emphasis of the combined cores is to provide center investigators with inter-project data useful for advancing the Center goals with a primary focus on data repositories that enable access, usability, integration and enhancement of the data.

Because the Conte Center teams collect vastly different data types, from text files to videos to neuroimaging data sets, both within and across species, we designed a lightweight, highly adaptable informatics platform with semantic capabilities to connect data across species. Further, because investigators across teams were unfamiliar with the raw and derived data generated by projects outside of their scientific domains, the system had to be capable of linking data, suggesting relevant data from the other project groups that could provide additional support for the within-project findings, while providing investigators with processed project data in a form useful for answering center-wide scientific questions. Given these requirements, the informatics group designed on a lightweight, graph-based information resource enabled by the Neuroimaging Data Model (NIDM)[3]. The main contributions of this work are in the application and evaluation of an information resource, powered by linked data techniques, and designed for sharing derived data from multiple species and across multiple domains in a dynamic geographically distributed research environment.

In the following sections we describe the components of the system, beginning with the overall architecture, followed by NIDM and the linked-data technologies used for developing the object models, a description of the web interface that facilitates query, visualization, and download of data from the graph-based database, a cross-species query example, and performance metrics.

## 2. Materials and Methods

In designing the Conte Center information resource, the informatics group initially evaluated the capabilities of each core project team in terms of their readiness to capture and store data associated with their project, the appropriateness of the data formats they planned on using with respect to sharing data with the other projects and cores, and an evaluation (in some cases, prediction) of which specific aspects of their data would be used by other Center projects. Example data sets were collected from each project and evaluated. The data sets comprised clinical, behavioral, and imaging data, both within and across species. Data were collected both from experimental and observational studies and comprise a range of variable types including categorical, ordinal, discrete and continuous. The wide range of data types suggested that a single data management solution capable of accommodating all raw/source and processed data from each project was infeasible. To test this intuition, we evaluated three publically available data management systems with the ability to capture many of the needed data types: the eXtensible Neuroimaging Archive Toolkit (XNAT)[4], the Human Imaging Database (HID)[5], and the Neuroinformatics Database (NiDB)[6]. These data management solutions were evaluated in terms of ease of data import, appropriateness of the data models for the complex multi-species data collected by the Center, intuitiveness of data storage and retrieval, and the functionality for querying across the diverse data types required. It was determined that although each solution was well suited for their respective use-cases, none was entirely appropriate for our intended use.

In evaluating intra-project data readiness, it was determined that each project had unique methods of capturing and storing their data and were comfortable accessing and analyzing these intra-project data. Further, after much consultation with project leaders and members of the BCDM core, it was determined that the majority of inter-project data sharing would occur on dynamic, processed data, rather than on source (raw) data directly. Further, each project had different preferences regarding the user interface to access the data, thus suggesting a set of domain-appropriate interfaces for query and retrieval. In aggregate, these observations led to a novel approach for the Conte Center informatics system, namely, a system that could store and share processed data, could accommodate dynamic data with flexible data models, and would facilitate sophisticated queries, linking data associated with different vocabularies across scientific domains.

### 2.1. Architecture Overview

Based on the observations and data collected during the data evaluation period, the high-level architecture for sharing processed data across project cores is shown in Figure 1. Each project team thus maintains their data internally (locally) in whatever form works best for their particular needs. Processed data, which are intended for combination and integration across project teams to advance the overall Center aims, are uploaded through a PHP web interface for entry by the informatics team into the multimodality cross-species data management system. For each data type an object model (see section 2.3 Object Models) is designed to model the data and associated semantics, using linked-data techniques and NIDM. Extract and transform modules (written in Python) process the source data according to the linked-data object models and save them as Resource Description Framework (RDF)[7] graphs using the Terse RDF Triple Language (TURTLE)[8]. The TURTLE documents are directly imported into the Virtuoso[9] database. Users then interact with the system through a PHP web interface with custom interfaces and linked-data query modules using the SPARQL[10] query language.

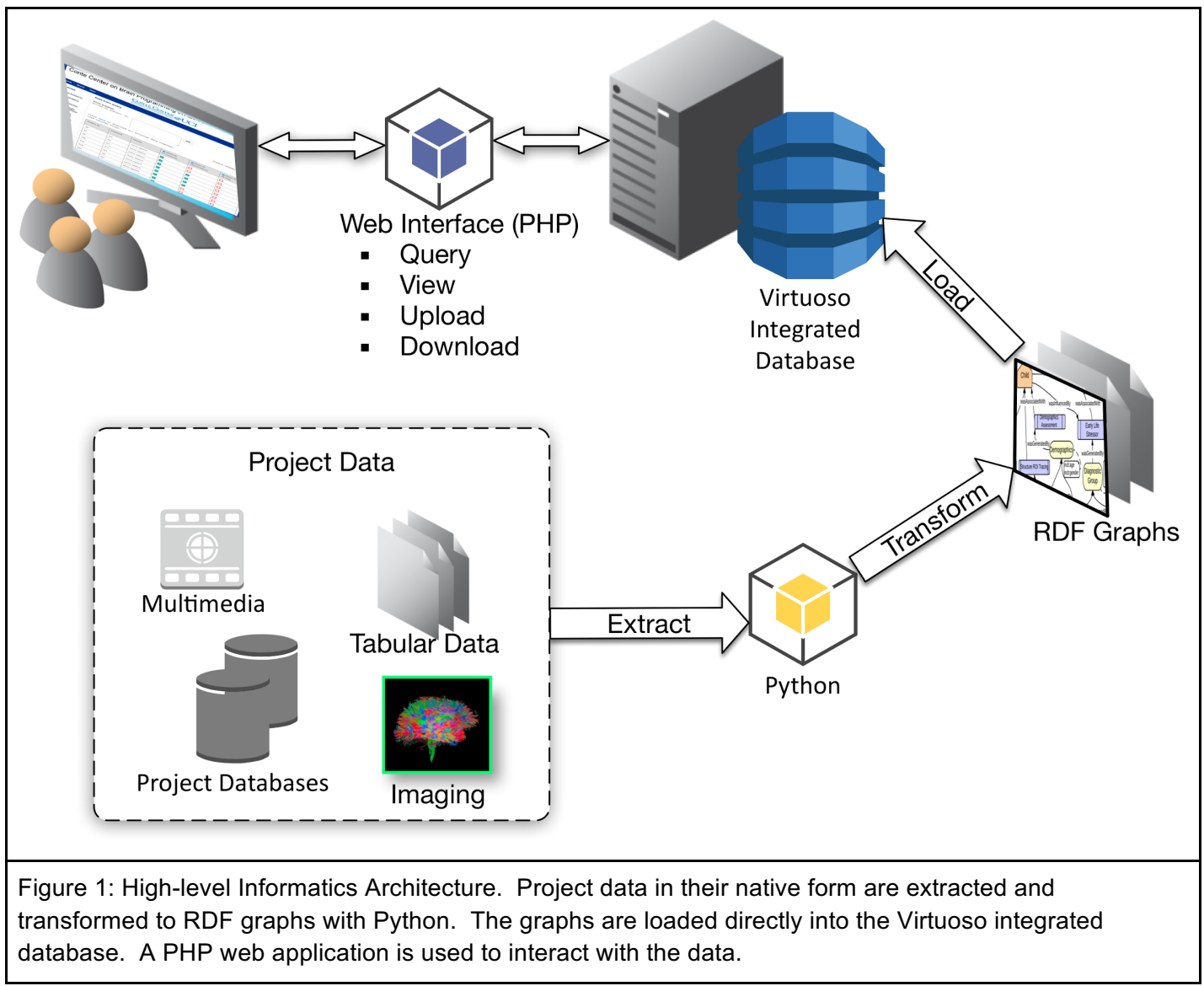

Figure 1: High-level Informatics Architecture. Project data in their native form are extracted and transformed to RDF graphs with Python. The graphs are loaded directly into the Virtuoso integrated database. A PHP web application is used to interact with the data.

## 2.2. Neuroimaging Data Model (NIDM)

NIDM was developed by an international team of cognitive scientists, computer scientists, and statisticians to create a data format capable of describing all aspects of the data lifecycle, from raw data through analyses and provenance. An early version used the *XML*-based Clinical and Experimental Data Exchange (XCEDE) XML format[11–13] developed under the Biomedical Informatics Research Network (BIRN) project[14]. Members of the BIRN project joined together with collaborators from the International Neuroinformatics Coordinating Facility (INCF) Data Sharing Task Force (NIDASH)[15] and began working on NIDM. The project adopted the Resource Description Framework (RDF) semantic web standards because of their extensibility, support for semantic annotation, and rich query language. NIDM was built on top of the PROV family of documents[16,17] and consists of three main interconnected specifications: Experiment, Results, and Workflows. These specifications and associated object models were envisioned, using a single core technology, to capture all aspects of the neuroimaging data lifecycle and to provide a critical component required for reproducibility and replication of research findings in the field, and for increasing the discoverability of data in shared resources. Although the components of NIDM were initially designed for neuroimaging, the processes used are extensible, i.e., modeling source data with object models, semantically tagging data with unambiguous terms, and transforming the source data into graph-based RDF documents. These processes and core NIDM vocabularies were used in the Conte Center resource wherever possible. For the unique data types in our project, custom object models were created, following the NIDM processes.

## 2.3. Object Models

An initial step in developing a linked-data representation is to create conceptual graphs or object models composed of the sets of information objects needed to represent the metadata that will ultimately be instantiated into an RDF document. Object models are abstractions that help people communicate about the organization of information using descriptive attributes that are required to unambiguously describe data. Building upon semantic web technologies, NIDM, and the PROV[17,18] family of documents, information in our application is organized into three main object representations: (1) *entities* represent physical, digital, and conceptual objects in the world, and are static; (2) *activities* represent dynamic processes that occur over a period of time and act upon or with entities; (3) *agents* are entities that have agency, i.e., can be responsible for an activity taking place. Using these core building blocks, a fixed set of relationships and descriptive attributes composed of Uniform Resource Identifiers (URI)[19] describing the precise meaning of the attributes as defined in formal ontologies and/or terminologies (e.g., National Cancer Institute Thesaurus), an unambiguous informational record of the data can be constructed.

For the Conte Center information resource we have developed six custom object models for rodent and human data and reused three existing models from NIDM-Experiment[20], NIDM FreeSurfer export[21], and the Connectome File Format[22]. To develop the custom models, the informatics team collaborated closely with each project team to generate the data representations that properly associate terms and relationships with each information type. Further, we collaborated with the NIDASH NIDM working group to assure that the models followed their basic design principles[23]. Figure 2 shows a simplified object model designed to describe structural and diffusion tensor imaging (DTI) segmentations from magnetic resonance imaging (MRI) scans collected on rodents. The model is composed of entities, activities, and agents as described (above) from the PROV data model, each with a varying number of additional attributes formatted as "namespace:term" qualified names, where namespace is a URI describing the terminology source and the term is an identifier describing the intended meaning of information artifact. In this example we use terms from the RDF vocabulary[24], the National Cancer Institute Thesaurus (NCIT)[25], the PROV data model, and terms specifically defined for the Conte Center at UC Irvine because they were unavailable in existing terminologies with appropriate definitions. The relationships among objects correspond to named associations from the PROV vocabulary, describing how the various objects interact to form a process graph for the underlying data. The choice of terms allows us to dynamically connect data at query time and develop interesting queries across data collected by different project teams.

Once the object models are fully specified, custom parsers are written in Python using the prov toolbox[26] to convert the source data into RDF documents according to the object models. For each object model we develop a separate parser. The RDF documents are then directly imported into the Virtuoso database and available for the application interface.

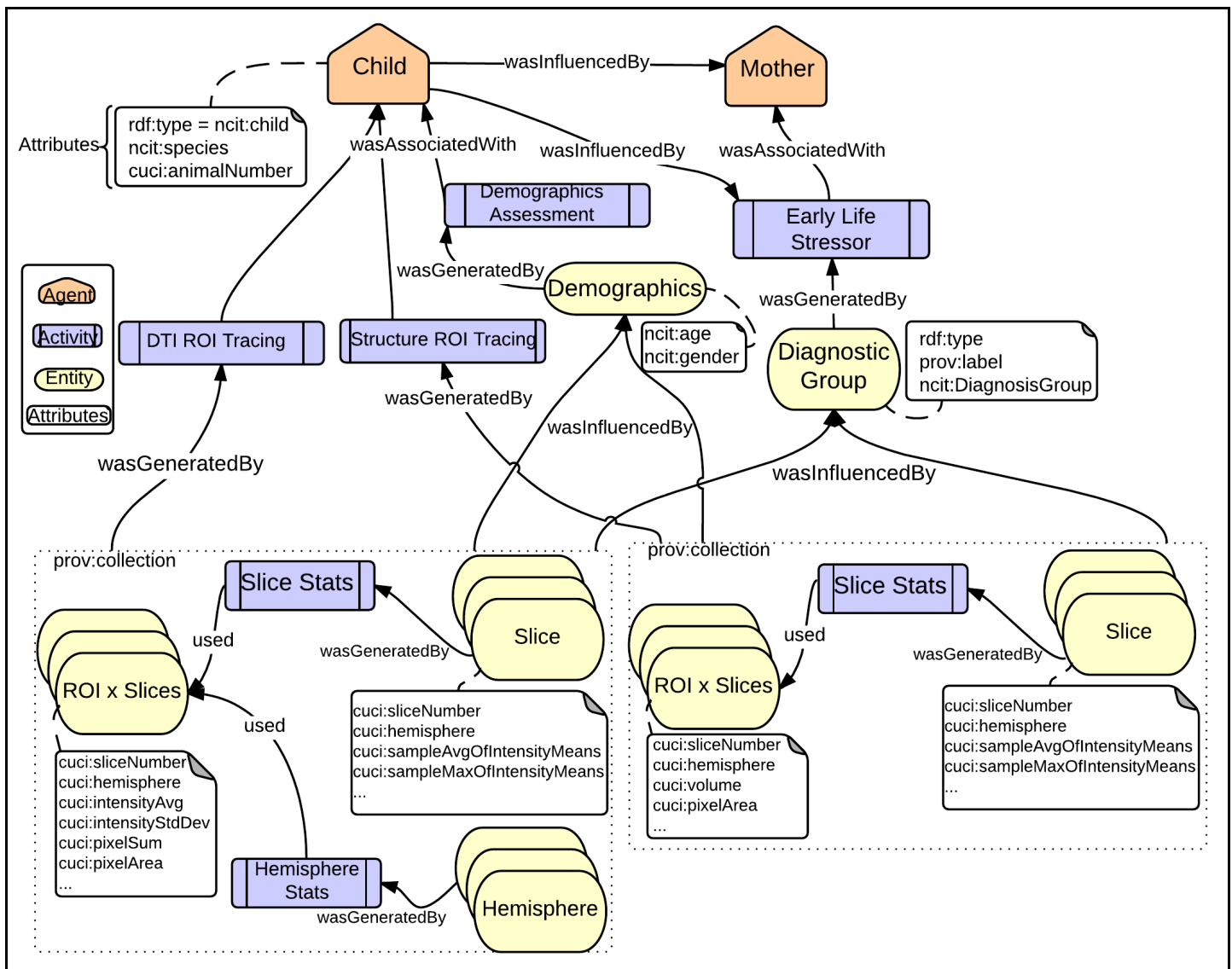

Figure 2: Object Model. Simplified rodent neuroimaging object model showing the relationships among the animal "Agents", the experimental condition (Early Life Stressor), simple demographics, and the region of interest (ROI) tracing statistics for both DTI and structural MRIs by slice (ROIxSlice) and combined by Slice and Hemisphere. For brevity, only a subset of attributes is shown.

### 2.4. System Architecture and Interface

The Conte Center informatics system was build using a standard three-tiered architecture consisting of user interface (UI), controller, and model/data access layers. The UI layer was built using a combination of PHP[27], Javascript[28], Dojo[29], HTML[30], and CSS[31] technologies. The UI layer provides interface handles for interaction with different parts of the application. An XMLHttpRequest object is created using Javascript when an event occurs at the UI. It sends asynchronous AJAX calls to controllers in the second tier and receives response data back, updating the UI. The Ajax Javascript framework allows the user interface to be updated without a full-page refresh. The controller layer was built using PHP and routes requests for data to the appropriate model interface in the data access layer, consistent with the user's query. After data have been received, the controller formats the data and routes them to the UI.

User interfaces were developed to query each object model with a simple and intuitive query menu on the left side of the interface shown in Figure 3 (top left panel). Organizing query/retrieval interfaces by data type provides a more intuitive interface for scientists to retrieve data from their domain. To assist in retrieving data across domains, we have built a "Show Subjects" interface, which provides all available data across project teams and species and have added cross-species queries to some interfaces (see section 3 Results and Discussion). Downloading data from this interface is facilitated through simple check boxes. Each query interface has functionality to display the underlying SPARQL query used to return the data. This utility was added to make it easier to learn how to query the system and to provide templated queries that can be modified to filter the data. Because each information type is tagged with a term and definition while transforming the source data to the RDF representation, adding "tool-tips" with variable definitions to the interface to guide domain scientists in constructing appropriate queries for their use-cases requires little additional work. This functionality is instrumental in providing researchers with real-time semantic information when they are using the system to download data from outside of their domain (bottom right panel of Figure 3). For imaging data, we have incorporated an XTK-enabled[32] UI, which allows users to interact with MRI images, structural connectomes, and functional image analysis results. This functionality has proven useful in allowing investigators to visually interrogate imaging data sets before downloading them for further modeling. Besides querying and viewing the data in the browser, users can download the text data as well as imaging data based on different selection criteria. Due to the large size of the data, queries are optimized to improve the performance and user experience (see section 3 Results and Discussion for query performance metrics).

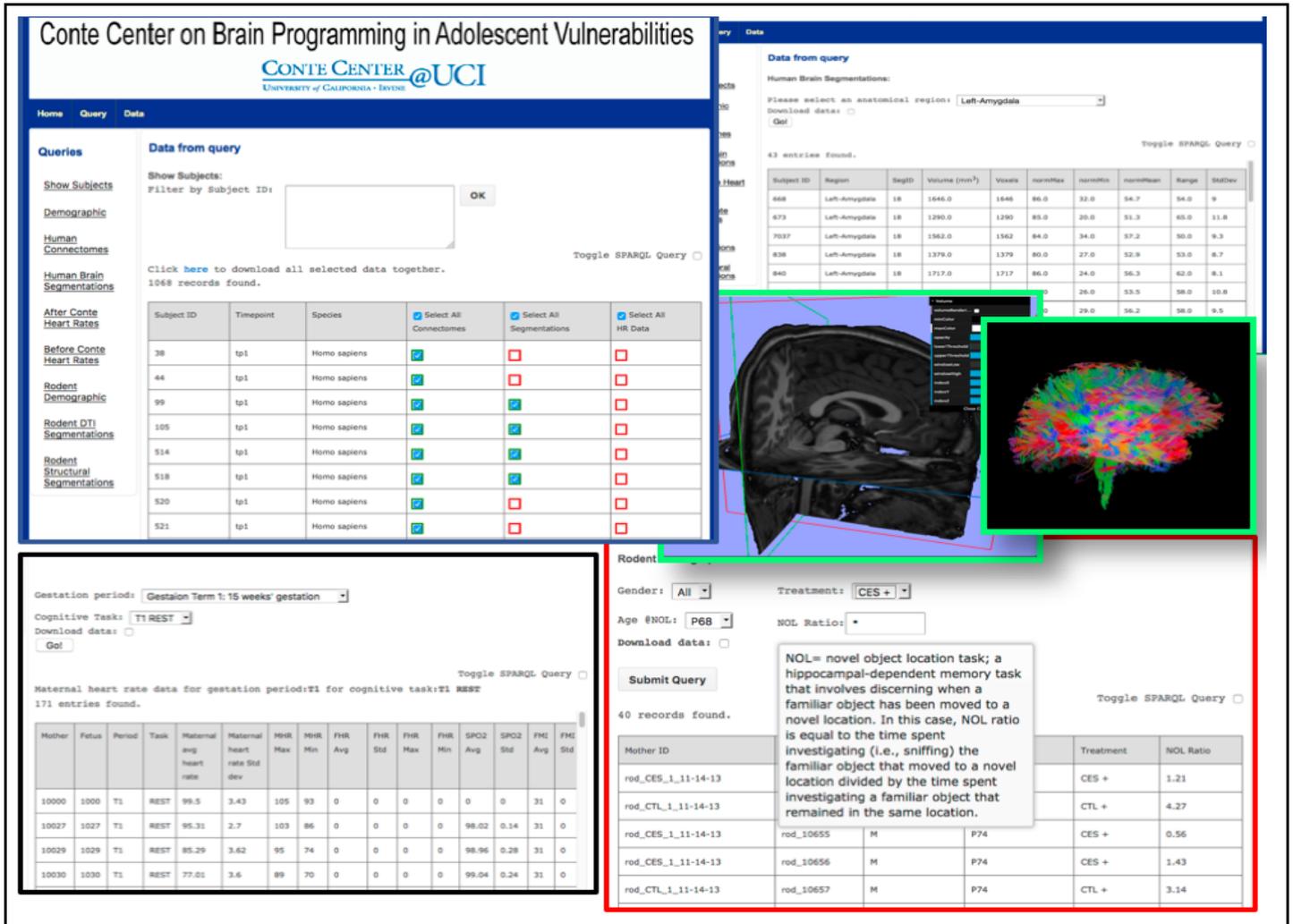

Figure 3: User Interface. Montage of selected user interface components of the system. Top left panel displays the union of all data in the database across species, and provides a simple selection system for downloading selected data types. The bottom left panel shows the interface for heart rate recording data,

where summary statistics are computed on the fly during query time. The top right panel shows the structural segmentation results with the XTK-enabled UI for viewing data. The bottom right panel shows information about rodents and experimental conditions.

3. **Results and Discussion**

The Conte Center informatics system has been useful in providing data to Center investigators, across project teams, in service of the overall specific aims of the Center. Users from each project team have been able to use the interface to find and download relevant data in a form useful for their particular analyses. The modular design, with object models and parsers/transformation modules for each unique data type that are linked to unique user interfaces, has resulted in an application that is easy to maintain and modify when source data formats change (e.g., when a data collection tool is updated and results in changes to the data format) and easy to extend by either adding additional object models or extending existing object models through links to separate sub-graphs. Further, the start-up development time and costs of such a custom application in support of a large multi-site center has been modest, consisting of a part-time programmer for one year.

Beyond providing interfaces to access data across project teams, we have begun implementing simple mapping functions to provide investigators with data across species. If a user selects to query the resource for human brain segmentation data, for example, the resource identifies that segmentation data is available in the rodent cohort of equivalent ages based on simple mapping functions. Our initial implementation consists of gross mappings between human and rodent ages. Such mappings are not unambiguous or uncontroversial, and thus our interface provides the user with the option of selecting the mapping function. Nevertheless, we have included a set of default linear functions for gross age mappings based on the experience of Conte Center investigators. To map from human to rodent ages, we use the following functions where the units of RodentAge are in postnatal days and those of HumanAge in postnatal years:

$$RodentAge(HumanAge) = \begin{cases} 7.5 + 2.1 * HumanAge & \text{if } HumanAge \geq 0 \\ 0 & \text{otherwise.} \end{cases}$$

$$HumanAge(RodentAge) = \begin{cases} -3.5 + 0.5 * RodentAge & \text{if } RodentAge \geq 7 \\ 0 & \text{otherwise.} \end{cases}$$

For example, in our user interface, if an investigator queries for rodent DTI segmentation data, our application, using the object model in Figure 2 and the SPARQL query in Table 1 (lines 1-5) retrieves the rodent ages corresponding to each DTI segmentation and computes an equivalent human age given the functions provided above. This query can be constructed by traversing the object model (Figure 2). Lines 1-3 of Table 1 query the attributes of the "Child" agent shown in the top of Figure 2. Next, lines 4-5 traverse the annotated links prov:wasGeneratedBy and prov:wasAssociatedWith to the "Demographics" entity which contains the rodent age attribute. We then bind a new variable, equivalent human age (equiv_human_age) to the result of the simple linear transformation (line 6). Using this mapped age, we can perform a similar query of the human data (lines 8-13), retrieving ages from the human cohort and filtering those children whose age matches those of the rodent ages (line 14).

| Line | SPARQL Query |
|------|--------------|
| 1 | ?rod_agent a prov:Agent ; |
| 2 |     ncit:species "Sprague-Dawley" ; |
| 3 |     cuci:animalNumber ?rodent_id . |
| 4 | ?demo_entity prov:wasGeneratedBy/prov:wasAssociatedWith ?rod_agent ; |
| 5 |     ncit:age ?rodent_age . |
| 6 | BIND(IF(?rodent_age >= 7,(-3.5 + 0.5*?rodent_age),0) as ?equiv_human_age) |
| 8 | ?agent_uri rdf:type prov:Agent ; |
| 9 |     ncit:species "Homo sapiens" ; |
| 10 |     ncit:subjectID ?child_id . |
| 11 | ?activity_uri prov:wasAssociatedWith ?agent_uri . |
| 12 | ?entity prov:wasGeneratedBy ?activity_uri ; |
| 13 |     ncit:age ?child_age . |
| 14 | filter((?child_age = ?equiv_human_age) |

Table 1: SPARQL query to find human children of comparable age to rodents using a linear age mapping.

We are working on incorporating more complex age-based maps of brain development across species such as those discussed in Avishai-Eliner et al.[34]. Ultimately we envision the resource to be flexible enough to answer more complex queries. For example, a researcher who performs structural analysis of high resolution T1-weighted 9.4 Tesla MRI scans might find that postnatal day 30-40 rodents raised in a fragmented environment (treatment CES+ in bottom-right panel of Figure 3) have smaller hippocampi bilaterally (data stored according to object model in Figure 2) than those reared in a more predictable environment. The Conte Center information resource would then suggest that there are human data of an analogous age range (e.g., 11-16 years old based on application of a properly cited mapping function) in a region related to the rodent hippocampus (based on application of a properly cited anatomical ontology). To facilitate similar queries, using our existing resource, we must have scientifically meaningful anatomical and age mappings across species. Although there is much work to be done to characterize the structural and functional homologies of neuroanatomical structures across rodent and human species, we are evaluating use of the Scalable Brain Atlas[35] and its web-based plugins for finding equivalent anatomical structures across species. With the appropriate anatomical mappings, we can augment the graphs during query time with relationships between structures allowing us to retrieve relevant linked data.

To evaluate the query performance of the system, we performed six typical queries, each repeated ten times, at different times of the day (Table 2). The queries were performed on the system hosting the Virtuoso database, containing an Intel Xeon E5-2609 1.7 GHz, eight-core CPU with 32GB of DDR4 ECC RAM. The queries were chosen to provide broad comparisons between large and small graphs and return sets. Overall the query times were not exceedingly long, the average query times ranging from approximately 0.5 seconds to 8 seconds. We carried out a statistical analysis to investigate the effects of graph size and return set size on query time. We worked with the logarithms of all variables to account for the wide range of problem sizes used in our experiments and to allow for the likelihood of power law relating the variables. Both graph size and return time are correlated with query time (Pearson's r for graph size and query time (on the log scale) is 0.80, Pearson's r for return set size and query time (on the log scale) is 0.59). The two size variables are themselves highly correlated (Pearson's r on the log scale is 0.86). A fixed-effects linear model indicates that the pair of variables together have significant explanatory power. The model resulted in a significant overall fit ($F=57.8$, $df=[2,57]$, $p<.0001$, Adj $R^2=0.66$) with both predictors contributing significantly to the model. We found that optimizing the SPARQL queries to reduce the number of intermediate triples improved query performance.

| Query | Type | Graph Size (triples) | Return Set (triples) | MeanTime (ms) | SD Time (ms) |
|---|---|---|---|---|---|
| 1 | Rodent Demographics | 1564 | 40 | 422.6 | 7.4 |
| 2 | Human Heart Rate,1 time point | 13299 | 390 | 1364.6 | 24.7 |
| 3 | Human Connectome | 20253 | 2451 | 645.2 | 11.2 |
| 4 | Human Heart Rate, 4 time points | 54499 | 863 | 7539.8 | 160.4 |
| 5 | Rodent Structural Segmentation | 96210 | 3674 | 2739.5 | 96.2 |
| 6 | Human Structural Segmentation | 1577291 | 4943 | 5535.8 | 57.1 |

Table 2: Mean (+/- SD) query times in milliseconds (ms) of ten repeated queries across differing graph complexities and return sizes.

## 4. Conclusions

The Conte Center at UC Irvine aims to understand the role of patterns of sensory inputs on the developing brain, focusing on their unpredictability and fragmentation, on long-term emotional and cognitive outcomes. The Program incorporates one key project focusing on rodent models and three others that focus on human behavior and biology. The animal data are used to generate hypotheses that can be tested in humans, and provide mechanistic explanations at the cellular, molecular, genetic, and epigenetic levels for phenomena that are measured in human and rodent at the behavioral and systems neuroscience levels.

This cross-species work generates substantial amounts of data that are of vastly different types, and have complex inter-relationships, some of which are known and others of which are speculative (e.g., the age and structure mappings mentioned above). Furthermore, these data come from different disciplines, with their own idiosyncratic vocabularies, and different physical locations. Our goal was to develop a data management system capable of handling such diverse types of data, their known and postulated relationships, with interfaces that are intuitive to researchers across all participating disciplines, and that supply data in a reusable and easily understood way across members of these same groups.

Our resulting system successfully manages the data from the cross-species, multi-center, integrated projects of the Conte Center at UC Irvine. A critical finding in assessing the Center informatics needs was that the principal informatics requirement was for data sharing across projects, including those from different disciplines and across species, and that within projects, locally managed idiosyncratic data storage and representations were preferred. Thus, our management approach focused on processed data, but nevertheless incorporated a wide variety of data types, from categorical data to text files to video to many different forms of brain imaging and neurophysiological data. For multi-species inferences, we are implementing a simple mapping mechanism, with default settings based on published research. To test these capabilities, we built a mapping function for rodent and human ages, and tested it on a number of typical queries at different times, with overall good performance. In summary, our informatics system, enabled by the Neuroimaging Data Model and powered by linked data techniques, permits data sharing across species and domains in a dynamic geographically distributed research environment.

Although sharing of the production information resource and data are not possible at this time because the project is ongoing, a Vagrant box of an earlier, modified, version of the information resource with example data from the NIDM working group has been made available for training purposes on GitHub (https://github.com/incf-nidash/nidm-training/tree/master/nidm_virtuoso_demo). Developers interested in the overall application can download the sample resource from GitHub and learn how the components were built to provide query and visualization using linked data.

**Acknowledgements:**
This work was supported by the National Institute of Mental Health of the National Institutes of Health under grant P50 MH 096889, "Fragmented early life environment and emotional / cognitive vulnerabilities" (Tallie Z. Baram, Center PI; Steven L. Small, Imaging Core PI, Hal Stern, BCDM Core PI). Their support is gratefully acknowledged.

**Competing Interests:** None

**References**

1. Home-new - Conte Center on Brain Programming in Adolescent Vulnerabilities. *Conte Center on Brain Programming in Adolescent Vulnerabilities* Available at: http://contecenter.uci.edu/. (Accessed: 5th April 2017)
2. Baram, T. Z. *et al.* Fragmentation and unpredictability of early-life experience in mental disorders. *Am. J. Psychiatry* **169,** 907–915 (2012).
3. Keator, D. B. *et al.* Towards structured sharing of raw and derived neuroimaging data across existing resources. *Neuroimage* **82,** 647–661 (2013).
4. Marcus, D. S., Olsen, T. R., Ramaratnam, M. & Buckner, R. L. The Extensible Neuroimaging Archive Toolkit: an informatics platform for managing, exploring, and sharing neuroimaging data. *Neuroinformatics* **5,** 11–34 (2007).
5. Ozyurt, I. B. *et al.* Federated web-accessible clinical data management within an extensible neuroimaging database. *Neuroinformatics* **8,** 231–249 (2010).
6. Book, G. A. *et al.* Neuroinformatics Database (NiDB) – A Modular, Portable Database for the Storage, Analysis, and Sharing of Neuroimaging Data. *Neuroinformatics* **11,** 495–505 (2013).
7. RDF - Semantic Web Standards. Available at: https://www.w3.org/RDF/. (Accessed: 6th April 2017)
8. RDF 1.1 Turtle. Available at: https://www.w3.org/TR/turtle/. (Accessed: 31st October 2016)
9. OpenLink Virtuoso Home Page. (2013). Available at: https://virtuoso.openlinksw.com/. (Accessed: 6th April 2017)
10. Website. Available at: https://www.w3.org/TR/rdf-sparql-query/. (Accessed: 31st October 2016)
11. Keator, D. B. *et al.* A general XML schema and SPM toolbox for storage of neuro-imaging results and anatomical labels. *Neuroinformatics* **4,** 199–212 (2006).
12. Gadde, S. *et al.* XCEDE: an extensible schema for biomedical data. *Neuroinformatics* **10,** 19–32 (2012).
13. Extensible Markup Language (XML). Available at: https://www.w3.org/XML/. (Accessed: 31st October 2016)
14. Glover, G. H. *et al.* Function biomedical informatics research network recommendations for prospective multicenter functional MRI studies. *J. Magn. Reson. Imaging* **36,** 39–54 (2012).
15. incf-nidash. incf-nidash/nidm. *GitHub* Available at: https://github.com/incf-nidash/nidm. (Accessed: 31st October 2016)
16. PROV-Overview. Available at: https://www.w3.org/TR/prov-overview/. (Accessed: 31st October 2016)
17. Moreau, L. *et al.* Special Issue: The First Provenance Challenge. *Concurr. Comput.* **20,** 409–418 (2008).
18. PROV-DM: The PROV Data Model. Available at: https://www.w3.org/TR/2013/REC-prov-dm-20130430/. (Accessed: 31st October 2016)
19. An example HTML document. Available at: http://www.ietf.org/rfc/rfc2396.txt. (Accessed: 31st October 2016)
20. NIDM-Experiment (under development). Available at: http://nidm.nidash.org/specs/nidm-experiment_dev.html. (Accessed: 7th April 2017)
21. satra. satra/nidm-notebooks. *GitHub* Available at: https://github.com/satra/nidm-notebooks. (Accessed: 7th April 2017)
22. Connectome File Format Library — Connectome File Format Library v2.0 documentation. Available at: http://cmtk.org/cfflib/. (Accessed: 7th April 2017)
23. Neuroimaging Data Model Primer (NIDM-Primer). Available at: http://nidm.nidash.org/specs/nidm-primer.html. (Accessed: 7th April 2017)
24. [No title]. Available at: http://www.w3.org/1999/02/22-rdf-syntax-ns#. (Accessed: 7th April 2017)
25. NCI Thesaurus. Available at: https://ncit.nci.nih.gov/ncitbrowser/. (Accessed: 7th April 2017)
26. trungdong. trungdong/prov. *GitHub* Available at: https://github.com/trungdong/prov. (Accessed: 7th April 2017)
27. PHP 5 Tutorial. Available at: https://www.w3schools.com/php/. (Accessed: 3rd May 2017)

28. JavaScript. *JavaScript.com* Available at: https://www.javascript.com. (Accessed: 3rd May 2017)
29. Dojo Toolkit. Available at: https://dojotoolkit.org/. (Accessed: 3rd May 2017)
30. HTML Tutorial. Available at: https://www.w3schools.com/html/. (Accessed: 3rd May 2017)
31. CSS Tutorial. Available at: https://www.w3schools.com/css/. (Accessed: 3rd May 2017)
32. xtk. xtk/X. *GitHub* Available at: https://github.com/xtk/X. (Accessed: 7th April 2017)
33. Quinn, R. Comparing rat's to human's age: How old is my rat in people years? *Nutrition* **21,** 775–777 (2005).
34. Avishai-Eliner, S., Brunson, K. L., Sandman, C. A. & Baram, T. Z. Stressed-out, or in (utero)? *Trends Neurosci.* **25,** 518–524 (2002).
35. Bakker, R., Tiesinga, P. & Kötter, R. The Scalable Brain Atlas: Instant Web-Based Access to Public Brain Atlases and Related Content. *Neuroinformatics* **13,** 353–366 (2015).